\documentclass[superscriptaddress,
twocolumn,showpacs,preprintnumbers,amsmath,amssymb]{revtex4}
\usepackage{graphicx}
\usepackage{dcolumn}
\usepackage{bm}
\usepackage{latexsym}
\begin{document}
\title{Intra-cellular traffic: bio-molecular motors on filamentary tracks{\footnote{Based on the invited talk delivered by DC at the IUPAP International Conference STATPHYS23, Genoa (Italy), July, 2007.}}}
\author{Debashish Chowdhury{\footnote{E-mail: debch@iitk.ac.in 
(Corresponding author)}}}
\affiliation{Department of Physics, Indian Institute of Technology, Kanpur 208016, India}
\author{Aakash Basu}
\affiliation{Department of Physics, Indian Institute of Technology, Kanpur 208016, India}
\author{Ashok Garai} 
\affiliation{Department of Physics, Indian Institute of Technology, Kanpur 208016, India}
\author{Philip Greulich} 
\affiliation{Institute for Theoretical Physics, University of Cologne,
D-50937 K\"oln, Germany} 
\author{Katsuhiro Nishinari} 
\affiliation{ Department of Aeronautics and Astronautics,
Faculty of Engineering, University of Tokyo,
Hongo, Bunkyo-ku, Tokyo 113-8656, Japan} 
\author{Andreas Schadschneider}
\affiliation{Institute for Theoretical Physics, University of Cologne,
D-50937 K\"oln, Germany} 
\author{Tripti Tripathi} 
\affiliation{Department of Physics, Indian Institute of Technology, Kanpur 208016, India}
\date{\today}%
\begin{abstract}
Molecular motors are macromolecular complexes which use some form 
of input energy to perform mechanical work. The filamentary tracks, 
on which these motors move, are made of either proteins (e.g., 
microtubules) or nucleic acids (DNA or RNA). Often, many such motors 
move simultaneously on the same track and their collective properties 
have superficial similarities with vehicular traffic on highways.
The models we have developed provide ``unified'' description: in the 
low-density limit, a model captures the transport properties of a 
single motor while, at higher densities the same model accounts for 
the collective spatio-temporal organization of interacting motors. 
By drawing analogy with vehicular traffic, we have introduced novel 
quantities for characterizing the nature of the spatio-temporal 
organization of molecular motors on their tracks. We show how the 
traffic-like intracellular collective phenomena depend on the 
mechano-chemistry of the corresponding individual motors. 
\end{abstract}
\pacs{89.20.-a;89.75.-k}
\maketitle

\section{Introduction} 

Motility is the hallmark of life. A distinguishing feature of molecular 
motor transport in eukaryotic cells is that the motor proteins move on 
filamentary ``tracks'' \cite{howard,schliwa,fisher,hackney}. 
The tracks for motor proteins are made of either proteins or nucleic 
acids. Not all motor proteins carry molecular cargo. A common feature of 
all these motors is that these perform mechanical work by utilizing some 
other form of input energy and hence the name ``motor''. All the molecular 
motors we consider in this paper directly convert chemical energy into 
mechanical work.

During several biological processes many motors move simultaneously on 
the same track. The collective movement of the motors under such 
cicumstances strongly resemble vehicular traffic flow \cite{polrev,physica}. 
In this paper we present few models of molecular motor traffic which, 
unlike earlier works of other groups, also capture the  essential 
steps of the mechano-chemistry of individual motors. 

Our models of molecular motor traffic are biologically motivated 
extensions of the totally asymmetric simple exclusion process (TASEP) 
\cite{schuetz}.  A TASEP is defined in terms of ``rules'' for updating 
the states of the system: a particle can move forward (with probability 
$q$) by one lattice specing if, and only if, the target site is empty.  
Updating can be implemented either in parallel or in random-sequential 
manner; properties of the model depends on the updating scheme. For a 
finite system, either periodic boundary conditions (PBC) or open boundary 
conditions (OBC) can be imposed. When the boundaries are open, at 
every time step, a particle can enter the system (and occupy the site 
$j=1$) with probability $\alpha$, if the site $j=1$ is empty. Similarly, 
under OBC, a particle occupying the site $j=L$ can exit with probability 
$\beta$. TASEP under OBC shows an interesting phase diagram and is the 
prototype for so-called boundary-induced phase transitions \cite{krug}. 
To our knowledge, TASEP is the simplest model of a system of interacting 
self-propelled particles on a discrete lattice. It has been extended in 
several ways to formulate ``particle-hopping'' models for capturing 
various interesting aspects of vehicular traffic \cite{css}. 

Our aim is to analyze molecular motor traffic from the perspective of 
vehicular traffic. Therefore, let us first list some of the important 
quantities used in traffic science for characterizing traffic flow 
\cite{css}. In the ``particle-hopping'' models of vehicular traffic 
each vehicle is represented by a particle. The number of particles 
leaving a detector site per unit time is defined as the {\it flux} 
and the relation between the average flux and the number density of 
the particles is called the {\it fundamental diagram}. 

In the particle-hopping models of vehicular traffic \cite{css}, the 
time interval in between the departure of successive particles 
from a detector site is defined as the {\it time headway} (TH). 
Let us start our clock as soon as a particle (let us label it by the 
integer $j = 0$) just leaves the detector whose location is fixed.  
Suppose, the next $n$ particles leave the same detector at times 
$t_1, t_2,...t_n$, respectively. Then, the corresponding THs are 
$\tau_1 = t_1$, $\tau_2 = t_2-t_1$, $\tau_3 = t_3-t_2$,...,
$\tau_n = t_n - t_{n-1}$, respectively. The average flux (averaged 
over the time interval $t_n$) is $J = n/t_n$. Since 
$\sum_{j=1}^{n} \tau_{j} = t_n$, 
$\biggl(\sum_{j=1}^{n} \tau_{j}\biggr)/n = 1/J$, i.e., 
the {\it mean TH} is the inverse of the average flux; therefore, 
the distribution of the THs contains more detailed informations on 
the flow characteristics of traffic than what is available from 
average flux \cite{thchow1,thchow2}.

The average density-profile of the particles is yet another quantitative 
characteristic of TASEP and TASEP-like models. A more detailed 
characterization of the spatial-organization of the particles is 
possible in terms the distribution of the {\it distance-headways} (DHs) 
where DH between two successive particles is defined to be the number of 
empty sites in between them \cite{dhchow,dhas}. 

In our earlier papers on models of molecular motor traffic 
\cite{nosc,greulich,tripathi,basu1,basu2,garairibo}
we reported results on average flux under both PBC and OBC as well as 
the average density profiles and the phase diagrams under OBC. In a few 
cases \cite{tripathi,garairibo}, we have also reported distributions 
of TH under OBC. In this paper, for each example of molecular motor 
traffic, we briefly summarize the model and the fundamental diagram 
before presenting new results on the distributions of THs. We also 
briefly discuss our new observations on the distributions of DHs in 
these models. The TH distributions have important biological 
implications and have been discussed in detail elsewhere 
\cite{tripathi,garairibo}. Three examples of molecular motor traffic 
are considered in the sections \ref{sec-kif1a}, \ref{sec-rnap} and 
\ref{sec-ribo}. The DH distributions in the models of molecular traffic 
are discussed briefly in section \ref{sec-dhdist}. Our summary and 
outlook is presented in section \ref{sec-conclusion}.

\section{Traffic of kinesins on microtubule track}
\label{sec-kif1a}

Microtubules and filamentous actin serve as the tracks for cytoskeletal 
motors and both are made of proteins \cite{howard}. The members of the 
{\it kinesin} and {\it dynein} superfamilies of cytoskeletal motors move 
on microtubules (MT) whereas those of the {\it myosin} superfamily move on 
actin filaments. These motors run on chemical fuel in the sense that the 
mechanical energy required for their movement is supplied from the energy 
released by the hydrolysis of adenosine triphosphate (ATP); the products 
of the hydrolysis reaction being adenosine diphosphate (ADP) and an  
inorganic phosphate \cite{hackney}. 

Cytoskeletal motors can attach at any motor-binding site on a MT track 
and can also detach from the track \cite{howard}. This distinct feature 
of the cytoskeletal motors was captured in the older TASEP-type models 
of molecular motor traffic \cite{lipo1,liporev,frey1,freyrev,santen,popkov1} 
by adding Langmuir-like attachment and detachment processes to the updating 
rules of TASEP. More specifically, in those models, a particle is 
allowed not only to hop forward, but also to ``attach'' to any empty 
lattice site (with rate $\omega_{a}$), and ``detach'' from an occupied 
site (with rate $\omega_{d}$). In a well-defined regime of parameters, 
this model exhibits co-existence of high-density and a low-density regions 
which are separated by a domain wall (shock); this type of spatio-temporal 
organization is interpreted as a jam in molecular motor traffic. But, 
since most of the effects of each mechano-chemical cycle of a motor is 
captured in these models through a single effective hopping rate $q$, it 
is difficult to make a direct quantitative comparison with experimental 
data which depend on such chemical processes. 

We have focussed attention on a family of single-headed kinesins, called 
KIF1A. In order to develope a model for cytoskeletal motor traffic, which 
would not suffer from the limitations of the earlier TASEP-like models, 
we have incorporated the essential steps of the mechano-chemical cycle 
of individual KIF1A motors in our model. 
Each biochemical cycle of a KIF1A motor consists of a sequence of four 
states, namely, kinesin (K), kinesin bound with ATP (KT), kinesin bound 
with ADP and phosphate (KDP) and, finally, kinesin bound with only ADP (KD). 
The motor binds strongly to the MT track in both the states K and KT; 
the state KDP has very short life time and KD binds weakly to the track. 
Therefore, at each spatial location in our simplified model, a KIF1A is 
allowed to exist in one of the two distinct ``chemical'' states depending 
on whether it is bound strongly or weakly to the track; these two chemical 
states are denoted by the symbols $S$ and $W$, respectively. 

\begin{figure}[ht]
\begin{center}
\includegraphics[angle=-90,width=0.85\columnwidth]{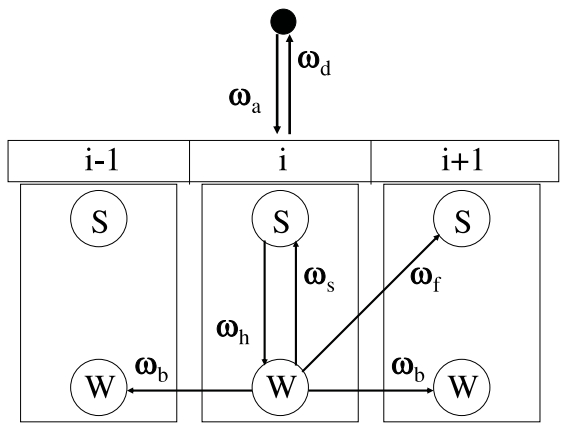}
\end{center}
\caption{Schematic description of the mechano-chemical cycle of a
single-headed kinesin motor KIF1A in our model. The equispaced 
sites labelled by the integers $...,i-1,i,i+1,...$ denote the 
binding sites of the motor on the microtubule (MT) track. The encircled 
symbols $S$ and $W$ denote the two ``chemical'' states of the 
motor in which it is, respectively, strongly and weakly bound to 
the track. The allowed transitions are indicated by the arrows and 
the symbols accompanying the arrows are the corresponding rate constants. 
}
\label{fig-kif1amod}
\end{figure}

The allowed transitions and the corresponding rate constants are shown in 
fig.\ref{fig-kif1amod}. The rate constants $\omega_{a}$ and $\omega_{d}$ 
account for the attachments and detachments of the motors. The rate 
constant $\omega_{b}$ corresponds to the unbiased one-dimensional 
Brownian motion of the motor in the state where it is weakly bound to 
the MT track. The rate constant $\omega_{h}$ is associated with the 
process driven by ATP hydrolysis which causes the transition of the 
motor from the state $S$ to the state $W$. The rate constants $\omega_{f}$ 
and $\omega_{s}$, together, capture the Brownian ratchet mechanism of 
movement of a KIF1A motor \cite{nosc,greulich}. As in the earlier 
TASEP-type models of cytoskeletal motor traffic
\cite{lipo1,liporev,frey1,freyrev,santen,popkov1}, 
none of the lattice sites is allowed to be occupied by more than one 
motor at a time.

Let $S_i(t)$ and $W_i(t)$ denote the probabilities of finding the motor 
at location $i$ in the strongly- and weakly-bound states, respectively.
The master equations for $S_i$ and $W_i$ are given by \cite{nosc}
\begin{eqnarray}
\frac{dS_i}{dt}&=& \omega_s W_i +\omega_f W_{i-1}(1-S_i-W_i) \nonumber\\
&&+\omega_a (1-S_i-W_i) -\omega_d S_i -\omega_h S_i \label{eqrc1}\\
\frac{dW_i}{dt}&=& \omega_h S_i + \omega_b (W_{i-1}+W_{i+1})(1-S_i-W_i) \nonumber\\
&&-\omega_s W_i - \omega_f W_i(1-S_{i+1}-W_{i+1}) \nonumber\\
&&-\omega_b W_i (2-S_{i+1}-W_{i+1}-S_{i-1}-W_{i-1}). \label{eqhc1}
\end{eqnarray}

Solving these equations in the steady state under PBC, we get the flux 
\begin{equation}
 J=\biggl[\frac{\omega_h}{\omega_h+\omega_s+\omega_f(1-\rho)}\biggr] \omega_f ~\rho ~(1-\rho).
  \label{fundall}
\end{equation}
where 
\begin{equation}
\rho = S + W = \frac{\Omega_h + \Omega_s + (\Omega_s+1)K 
- \sqrt{D} + 2}{2(1+K)}\, .
\end{equation}
is the steady-state number-density of the motors on the MT 
track, with $K=\omega_d/\omega_a$,
$\Omega_h=\omega_h/\omega_f$, $\Omega_s=\omega_s/\omega_f$, and
\begin{equation}
 D=4\Omega_s K(1+K)+
(\Omega_h +\Omega_s + ( \Omega_s-1)K)^2.
\end{equation}

\begin{figure}[ht]
\begin{center}
\includegraphics[width=0.9\columnwidth]{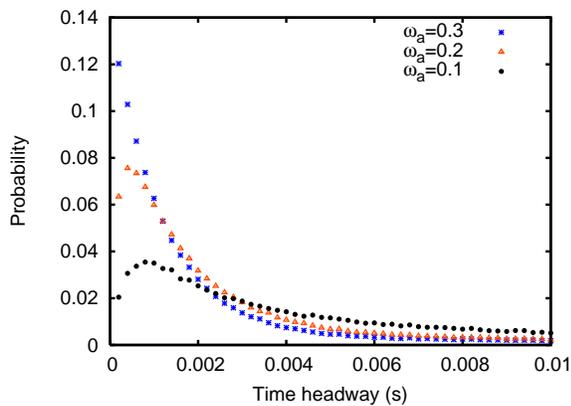}
\end{center}
\caption{TH distributions in the model of KIF1A traffic under OBC. The 
three curves correspond to three different values of $\omega_{a}$ (in 
the units of $s^{-1}$). The other parameters are $\omega_{d}=0.1 s^{-1}$, 
$\omega_{h}=100 s^{-1}$, $\omega_{s}=145 s^{-1}$, $\omega_{f}=55 s^{-1}$, 
$\omega_{b}=1125 s^{-1}$, $\alpha=50 s^{-1}$ and $\beta=700 s^{-1}$.}
\label{fig-kif1aTH}
\end{figure}

In this model, {\it departure} of a motor from a site can take place 
in two different ways: either by hopping to the next site or by 
detachment from the track. Therefore, we first modify the original 
definition of TH, which we presented in the introductory section, 
for our model of KIF1A under OBC. We interpret the attachments at 
the site $i=1$ and detachments at $i = L$ as consequences solely of 
forward hoppings of the motors, rather than manifestations of 
Langmuir-like kinetics. The time interval in between the exit of 
two successive motors from the site $i = L$ is identified as the 
corresponding TH. 

A few typical TH-distributions under OBC in our model of KIF1A traffic 
are plotted in fig.\ref{fig-kif1aTH} \cite{agthesis}. A higher 
$\omega_{a}$ leads to a higher average density of motors on the track 
which, in turn, reduces the most probable TH as long as $\beta$ remains 
sufficiently high. This trend of variation is consistent with that of 
the average flux with $\omega_{a}$ \cite{nosc,greulich}. Moreover, a 
wider distribution at smaller $\omega_{a}$ indicates stronger 
fluctuations in the THs at a lower average density. In principle, 
these new theoretical predictions can be tested by carrying out 
{\it in-vitro} experiments with fluorescently-labelled KIF1A molecules 
using single-molecule imaging techniques. Systematic study of the 
variation of the width of the TH distribution with the control 
parameters of the experiment will provide deep insight into the nature 
of ``noise'' in intracellular cytoskeletal motor transport.

\section{Traffic of RNAP motors on DNA tracks}
\label{sec-rnap}

Polymerization of a mRNA from the corresponding DNA template is carried 
out by a motor called RNA polymerase (RNAP) \cite{wangrev} and the 
process is called transcription. To our knowledge, all the models of 
transcription reported earlier 
\cite{julicherrnap98,osterrnap98,sousa96,sousa97,sousa06,mdwang04,mdwang07,nudler05,tadigotla,peskinrnap06,woo06}
capture only the stochastic mechano-chemistry of the individual RNAP
motors. Cooperation and collisions between RNAP motors is known to
have non-trivial effects on the rate of transcription
\cite{nudler03a,nudler03b,crampton06,sneppen05}. Moreover, the
possibility of the formation of queues in RNAP traffic has also been
explored \cite{bremer95}.  

\begin{figure}[ht]
\begin{center}
\includegraphics[angle=-90,width=0.85\columnwidth]{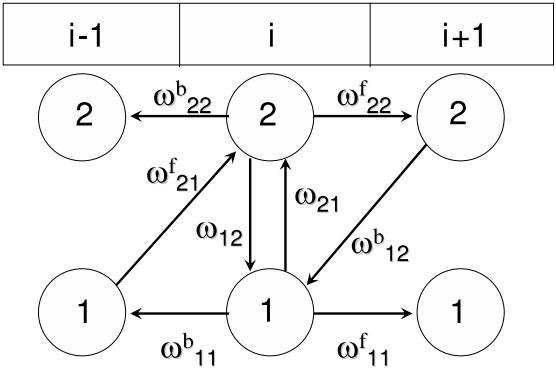}
\end{center}
\caption{A schematic representation of the mechano-chemical cycle of
each RNAP in our model. The equispaced sites labelled by the integers 
$...,i-1,i,i+1,...$ denote the nucleotides on the template DNA track. 
No $PP_{i}$ is bound to the RNAP in the state $1$ whereas the 
pyrophosphate ($PP_{i}$)-bound state of the RNAP is labelled by the 
index $2$. The allowed transitions denoted by arrows and the 
corresponding rate constants are also shown.
}
\label{fig-rnapmod}
\end{figure}

The interactions of RNAPs in transcriptional interference \cite{shearwin05}
is a well known phenomenon and it has also been modelled quantitatively
\cite{sneppen05}. However, instead of studying interactions of RNAPs
during the transcription of different genes, we have modelled the
steric interactions of RNAPs which are simultaneously involved in the
transcription of the same gene.

The model is described schematically in fig.\ref{fig-rnapmod}. Each of 
the lattice sites corresponds to a single nucleotide on the DNA template. 
Each successful addition of a nucleotide to the elongating RNA is 
accompanied by a forward stepping of the RNAP. A single mechano-chemical 
cycle of the RNAP during this elongation stage consists of several 
steps of which the major ones are as follows: (i) Nucleoside 
triphosphate (NTP) binding to the active site of the RNAP when the 
active site is located at the tip of the growing RNA transcript, 
(ii) NTP hydrolysis, (iii) release of pyrophosphate ($PP_i$), one 
of the products of hydrolysis, and (iv) accompanying forward 
stepping of the RNAP. Since $PP_i$-release is known to be the 
rate-limiting step, we consider only two distinct chemical states 
$\mu$ of the RNAP; $\mu =1$ refers to the state in which the RNAP is 
not bound to any $PP_{i}$ whereas $\mu = 2$ corresponds to the state 
with bound $PP_{i}$.

The processes corresponding to the rate constants $\omega_{12}$ and 
$\omega_{21}$ are dominated by $PP_{i}$-release and its reverse reaction. 
The symbol $\omega^{f}_{21}$ is the rate of the successful addition of 
an NTP, catalyzed by the RNAP, whereas $\omega^{b}_{12}$ is that of the 
reverse reaction. The remaining four rate constants correspond to 
polymerization/depolymerization of the RNA, by one monomer, unaided by 
the RNAP. Since premature detachment of a RNAP from its track is a very 
rare event during the elongation of the growing RNA, we do not 
allow such processes in our model.

\begin{figure}[ht]
\begin{center}
\includegraphics[width=0.9\columnwidth]{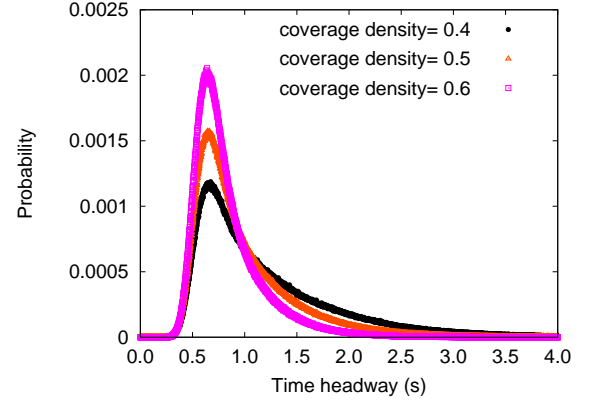}
\end{center}
\caption{Distributions of the time-headways in the model of RNAP 
traffic under PBC. The three curves correspond to three different 
coverage densities of the RNAP motors, all for fixed  
$\omega^{f}_{21} = 100$ and $\omega_{21} = 1.0$.
}
\label{fig-rnapTHd}
\end{figure}

\begin{figure}[ht]
\begin{center}
\includegraphics[width=0.9\columnwidth]{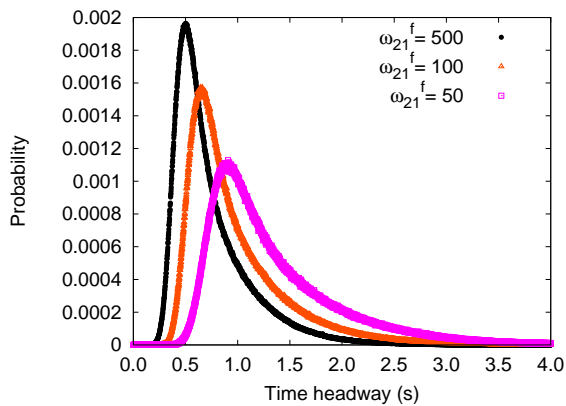}
\end{center}
\caption{Distributions of the time-headways in the model of RNAP 
traffic under PBC. Different curves correspond to different 
values of $\omega^{f}_{21}$, all for the fixed coverage density 
$0.5$ and $\omega_{21} = 1.0$.
}
\label{fig-rnapTHntp}
\end{figure}

\begin{figure}[ht]
\begin{center}
\includegraphics[width=0.95\columnwidth]{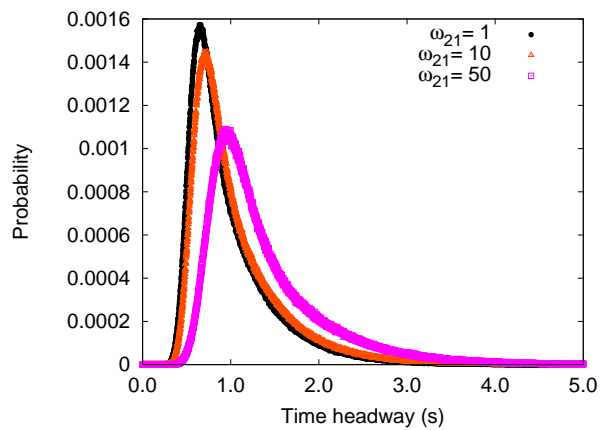}
\end{center}
\caption{Distributions of the time-headways in the model of RNAP
traffic under PBC. Different curves correspond to different 
values of $\omega_{21}$, all for the fixed coverage density $0.5$ 
and $\omega^{f}_{21} = 100$.
}
\label{fig-rnapTHppi}
\end{figure}

The values of some of the rate constants used in our numerical studies 
are as follows: $\omega_{12} = 31.4 ~s^{-1}$, 
$\omega^{b}_{12} = 0.21 ~s^{-1}$, 
$\omega^{f}_{11} = 4.66 \times 10^{-5} ~s^{-1}$,
$\omega^{b}_{11} = 9.4 ~s^{-1}$,  
$\omega^{f}_{22} = 0.31 \times 10^{-6} ~s^{-1}$,
$\omega^{b}_{22} = 0.063 ~s^{-1}$. The numerical values of the other 
rate constants are given in the captions of the figures 
\ref{fig-rnapTHd}, \ref{fig-rnapTHntp} and \ref{fig-rnapTHppi}.

Unlike the cytoskeletal motors, a single RNAP is so large that it 
can simultaneously cover $r$ successive nucleotides on the track.
In our terminology, a site is {\it occupied} by 
a RNAP if it coincides with the leftmost of the $r$ sites representing 
that RNAP while the next $r-1$ sites on its right are said to be 
{\it covered} by the same RNAP. Irrespective of the actual numerical 
value of $r$, each RNAP can move forward or backward by only one site 
in each time step, if demanded by its own mechano-chemistry, provided 
the target site is not already covered by any other RNAP.  
The total number of RNAPs on the DNA template is denoted by the 
symbol $N$. Thus, $\rho = N/L$ is the {\it number density} of the
RNAPs. The  {\it coverage density} is defined by $\rho_{cov} = N r/L$
which is the total fraction of the nucleotides covered by all the RNAPs
together.

Let $P_{\mu}(i,t)$ denote the probability that there is a RNAP at the
spatial position $i$ and in the chemical state $\mu$ at time $t$. 
Let $Q(\underline{i}|j)$ be the conditional probability that, given a 
RNAP at site $i$, site $j$ is empty. Note that, if site $i$ is given 
to be occupied by one RNAP, the site $i-1$ can be covered by another 
RNAP if, and only if, the site $i-r$ is also occupied. 
In the mean-field approximation, the master equations for $P_{\mu}(i,t)$
are given by \cite{tripathi}
\begin{eqnarray}
\frac{dP_{1}(i,t)}{dt} &=& ~\omega^{f}_{11} ~P_{1}(i-1,t) ~Q(\underline{i-1}|i-1+r) \nonumber \\
&+& ~\omega^{b}_{11} ~P_{1}(i+1,t) ~Q(i+1-r|\underline{i+1}) \nonumber \\
&+& ~\omega^{b}_{12} ~P_{2}(i+1,t) ~Q(i+1-r|\underline{i+1}) \nonumber \\
&+& ~\omega_{12} ~P_{2}(i,t) - ~\omega_{21} ~P_{1}(i,t) \nonumber \\
&-& ~(\omega^{f}_{11}+\omega^{f}_{21}) ~P_{1}(i,t) ~Q(\underline{i}|i+r) \nonumber \\ 
&-& ~\omega^{b}_{11} ~P_{1}(i,t) ~Q(i-r|\underline{i})
\label{eq-masterp1}
\end{eqnarray}
\\
\begin{eqnarray}
\frac{dP_{2}(i,t)}{dt} &=& ~\omega^{f}_{22} ~P_{2}(i-1,t) ~Q(\underline{i-1}|i-1+r) \nonumber \\
&+& ~\omega^{b}_{22} ~P_{2}(i+1,t) ~Q(i+1-r|\underline{i+1}) \nonumber \\
&+& ~\omega^{f}_{21} ~P_{1}(i-1,t) ~Q(\underline{i-1}|i-1+r) \nonumber\\
&+& ~\omega_{21} P_{1}(i,t) - ~\omega_{12} ~P_{2}(i,t) \nonumber \\
&-& ~(\omega^{b}_{22}+\omega^{b}_{12}) ~P_{2}(i,t) ~Q(i-r|\underline{i})\nonumber \\
&-& ~\omega^{f}_{22} ~P_{2}(i,t) ~Q(\underline{i}|i+r) 
\label{eq-masterp2}
\end{eqnarray}
In the steady state under PBC, 
\begin{eqnarray}
P_{1} = \biggl(\frac{\omega_{12} + \omega^{b}_{12}Q}{\Omega_{\updownarrow} + \Omega_{\leftrightarrow}Q}\biggr) ~\rho  \nonumber \\
P_{2} = \biggl(\frac{\omega_{21} + \omega^{f}_{21}Q}{\Omega_{\updownarrow} + \Omega_{\leftrightarrow}Q}\biggr) ~\rho
\label{eq-stsolnpbc}
\end{eqnarray}
where
\begin{eqnarray}
\Omega_{\updownarrow} = \omega_{12} + \omega_{21} \nonumber \\
\Omega_{\leftrightarrow} = \omega^{f}_{21} + \omega^{b}_{12} 
\end{eqnarray}
and $Q$ is given by
\begin{eqnarray}
Q(\underline{i}|i+r) = Q(i|\underline{i+r}) = \frac{1-\rho r}{1+\rho-\rho r}
\label{eq-condq2}
\end{eqnarray}
The corresponding steady-state flux is given by 
\begin{eqnarray}
J &=& ~\Omega_{1} ~P_{1} ~Q + ~\Omega_{2} ~P_{2} ~Q \nonumber \\
&=& (~\Omega_{1} ~P_{1} + ~\Omega_{2} ~P_{2}) \biggl(\frac{1-\rho_{cov}}{1+\rho-\rho_{cov}}\biggr) 
\label{eq-stfluxpbc}
\end{eqnarray}
where 
\begin{eqnarray} 
\Omega_{1} = \omega^{f}_{11} + \omega^{f}_{21} - \omega^{b}_{11} \nonumber \\
\Omega_{2} = \omega^{f}_{22} - \omega^{b}_{12} - \omega^{b}_{22}, 
\end{eqnarray}
are two {\it effective forward hopping rates} from the states $1$ and $2$,
respectively.

Since, unlike vehicles in highway traffic, an RNAP motor can step 
backward, TH under PBC must be computed carefully as follows 
\cite{ttthesis}: we label the motors sequentially in the beginning. 
Because of the PBC and because of the impossibility of overtaking, 
the sequence of the labels remain unchanged during the time evolution 
of the system. TH is defined as the time interval between the {\it 
departure} (i.e., forward hopping) of the {\it successive} motors 
from the same site.

Typical TH distributions of the RNAPs in this model under PBC are plotted 
in figs.\ref{fig-rnapTHd}, \ref{fig-rnapTHntp}, \ref{fig-rnapTHppi} for 
biologically relevant sets of values of the parameters \cite{ttthesis}; 
the corresponding results under OBC have been reported elsewhere 
\cite{tripathi}. In these parameter regimes, because of the ``chemical 
transition'' between the states $1$ and $2$, the minimum TH is non-zero, 
i.e., $P(\tau) = 0$ for all $\tau \leq \tau_{min}$. The magnitude of 
$\tau_{min}$ as well as the most probable TH are practically independent 
of the coverage density. Moreover, both $\tau_{min}$ and the most probable 
TH decrease with increasing $\omega^{f}_{21}$, (i.e., with increasing 
concentration of the monomeric subunits of the growing mRNA). In 
contrast, $\tau_{min}$ and the most probable TH increase with increasing 
$\omega_{21}$ which tends to suppress forward movement of each RNAP 
along the main pathway (see fig.\ref{fig-rnapmod}). The trends of 
variation of the most probable TH with the model parameters are 
consistent with the corresponding trends of variation of the average 
flux with the same parameters. 

Very recently we have proposed \cite{tripathi} that the width of the TH 
distribution, as defined above in the context of RNAP traffic, can serve  
as a good quantitative measure of noise in transcription. Recent progress 
in imaging techniques has made it possible to monitor the synthesis of 
successive individual RNAs in a living cell \cite{shavtal04}. Using these 
techniques it has been demonstrated \cite{golding05,chubb06,raj06}
that synthesis of RNAs take place in ``bursts'' \cite{golding2}. The 
time series of the events correspoding to the TH distributions shown in 
figs.\ref{fig-rnapTHd}, \ref{fig-rnapTHntp} and \ref{fig-rnapTHppi} do 
not account for such ``bursts''. But ``bursting'' is observed when our 
model is extended by incorporating the processes of ``swtiching on'' and 
``switching off'' of the gene \cite{tripathi2}.

\section{Traffic of ribosomes on mRNA tracks}
\label{sec-ribo}

Synthesis of a protein from the corresponding messenger RNA (mRNA) is 
carried out by a motor called ribosome \cite{spirinbook} and the process 
is referred to as translation of genetic code. Usually, many ribosomes 
move simultaneously on a single mRNA strand while each synthesizes a 
separate copy of the same protein. Strictly speaking, a ribosome is 
neither a particle nor a hard rod 
\cite{macdonald68,macdonald69,lakatos03,shaw03,shaw04a,shaw04b,chou03,chou04,dong}; 
its mechanical movement along the mRNA track is coupled to its internal 
mechanochemical processes which also drive the synthesis of the protein. 

\begin{figure}[ht]
\begin{center}
\includegraphics[angle=-90,width=0.95\columnwidth]{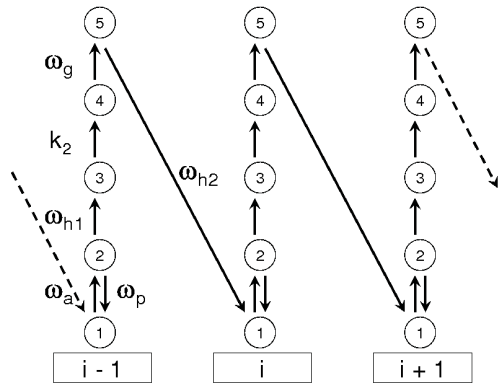}
\end{center}
\caption{A schematic representation of the biochemical cycle of a
single ribosome during the elongation stage of translation in our
model \cite{basu1}. The index below the box labels the codon 
on the mRNA with which the ribosome binds. Each circle labelled by 
an integer index represents a distinct state in the mechano-chemical 
state of a ribosome. The symbols accompanied by the arrows define 
the rate constants for the corresponding transitions from one state 
to another.
}
\label{fig-ribomod}
\end{figure}

Very recently, we have developed a model that not only incorporates the 
inter-ribosome steric interactions, but also captures explicitly the 
essential steps in the intraribosome chemomechanical processes. 
This model for ribosome traffic is very similar to that for RNAP traffic 
except that (i) each lattice represents a codon (a triplet of nucleotides) 
and (ii) the mechano-chemical of a single ribosome 
(see fig.\ref{fig-ribomod}) is quite different from that of a RNAP. 

Each ribosome consists of two subunits. The smaller subunit binds to 
the mRNA track. But, the actual elongation of the protein takes place 
in the larger subunit by the addition of the successive monomers,  
called amino acid. The operations of the two subunits are coordinated 
by an adaptor molecule called tRNA. It uses the anticodon at one of 
its ends to decode the genetic instructions stored in each codon of 
the mRNA which serves also as the template for synthesis of a protein. 
Each tRNA carries an amino acid at its other end. Correct codon-anticodon 
matching ensures that the correct amino acid, as dictated by the mRNA 
template, is used by the larger subunit for elongating the protein by 
one monomer.

The main steps in the mechano-chemical cycle of a ribosome are the 
following:\\
(i) arrival of the correct tRNA (rate constant $\omega_{a}$), 
(ii) growth of the protein by one monomer because of the formation 
of a covalent bond with the newly arrived amino acid (rate $\omega_{g}$), 
and 
(iii) forward movement of the ribosome by one codon which is associated 
with the hydrolysis of a guanosine triphosphate (GTP) molecule (rate 
$\omega_{h2}$). The rate constants $\omega_{h1}$ and $k_{2}$ are 
associated with two steps during which another GTP molecule is 
hydrolyzed. A more detailed description of the model is available in
ref.\cite{basu1}.
The biologically relevant values of the parameters, which have been kept 
fixed throughout our studies of the TH distributions, are as follows: 
$\omega_{p} = 0.0028 s^{-1}$, $k_{2} = 2.4 s^{-1}$ and 
$\omega_{g} = 2.5 s^{-1}$.

\begin{figure}[ht]
\begin{center}
\includegraphics[width=0.9\columnwidth]{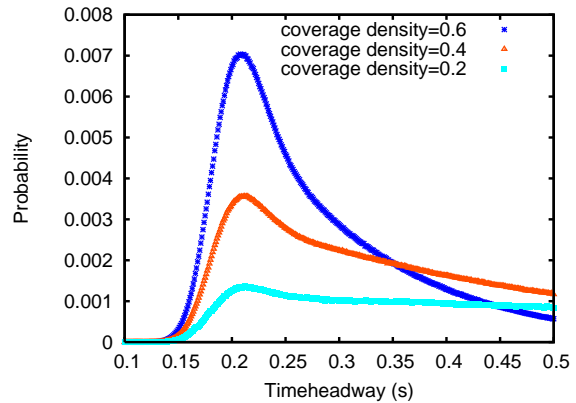}
\end{center}
\caption{TH distributions in the model of ribosome traffic under 
PBC. The three curves correspond to three different coverage 
densities of the ribosomes, all for $\omega_{a} = 2.5 s^{-1}$ 
and $\omega_{h1} = \omega_{h2} = \omega_{h} = 10 s^{-1}$. 
}
\label{fig-riboTHd}
\end{figure}

\begin{figure}[ht]
\begin{center}
\includegraphics[width=0.9\columnwidth]{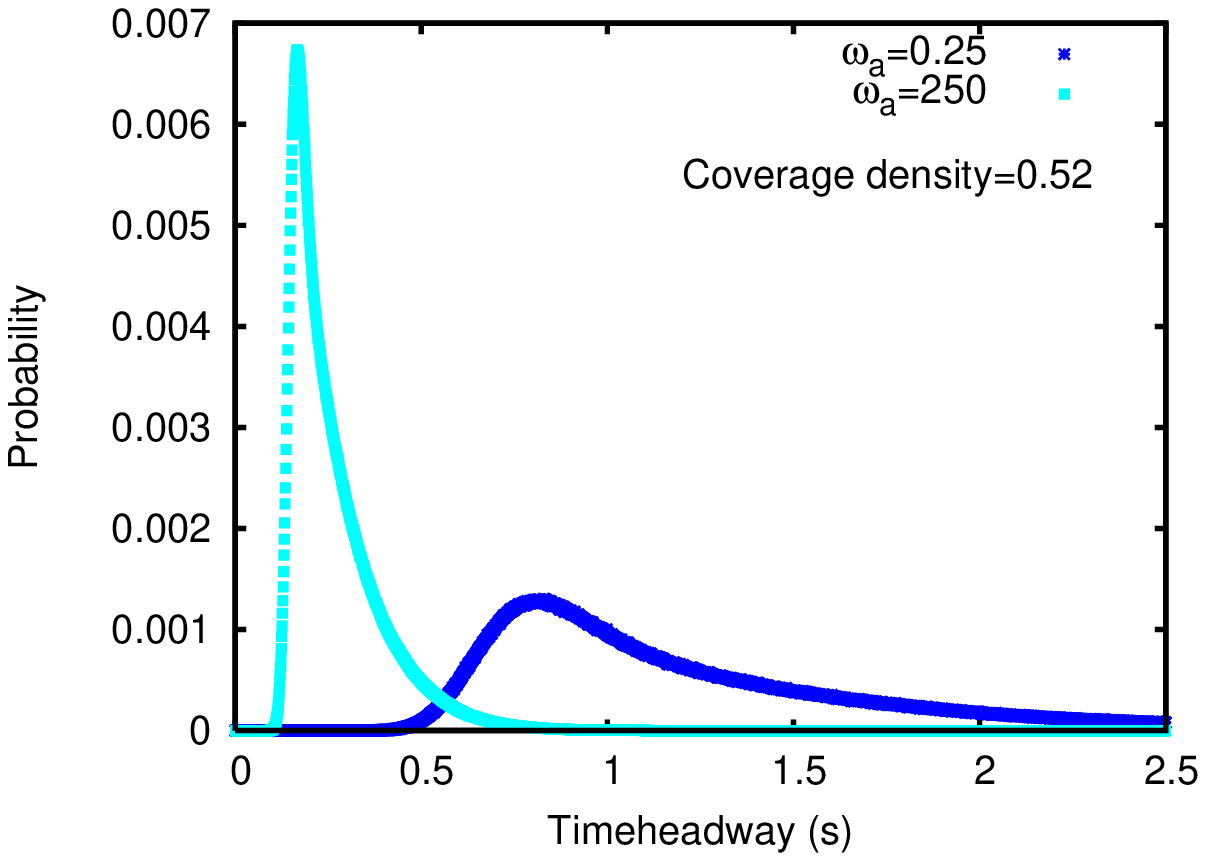}
\end{center}
\caption{TH distributions in the model of ribosome traffic under 
PBC. Different curves correspond to different values of $\omega_{a}$, 
all for coverage density = 0.52 and 
$\omega_{h1} = \omega_{h2} = \omega_{h} = 10 s^{-1}$. 
}
\label{fig-riboTHwa}
\end{figure}

\begin{figure}[ht]
\begin{center}
\includegraphics[width=0.9\columnwidth]{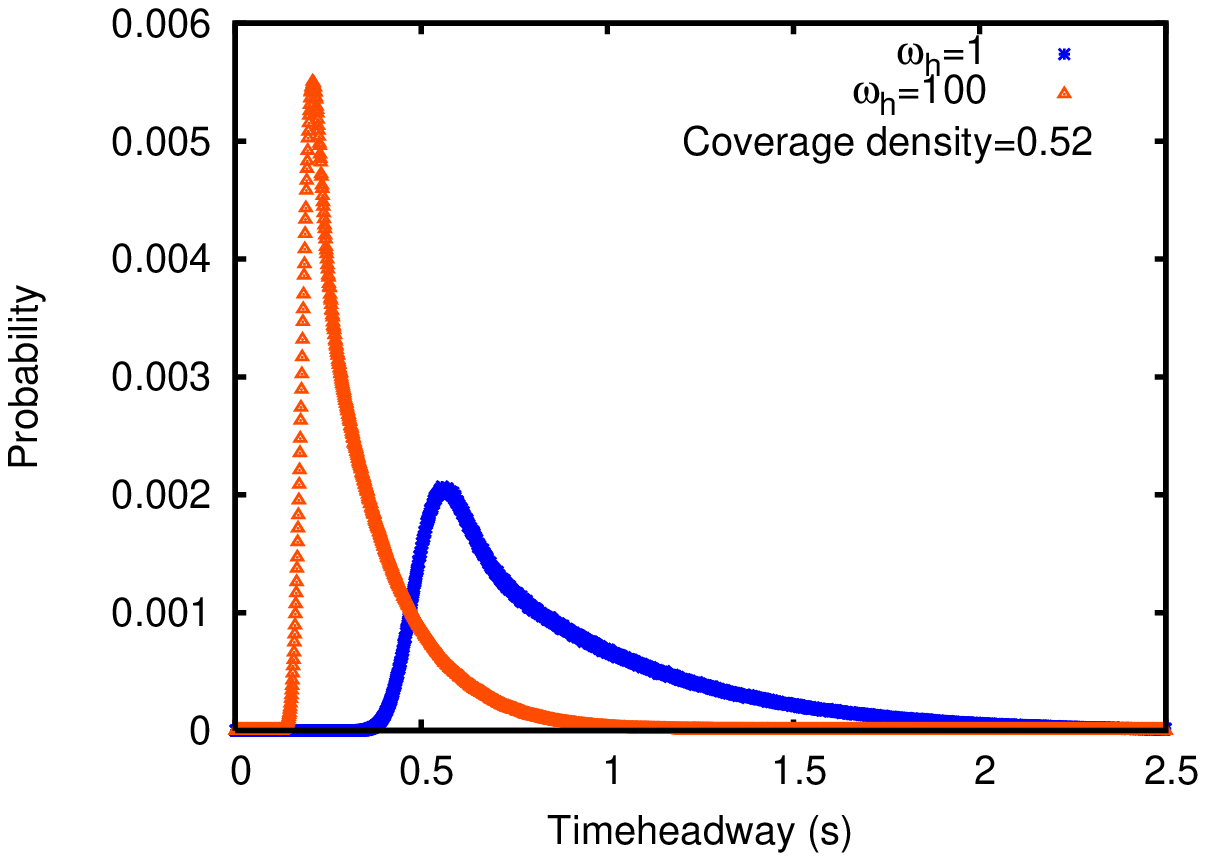}
\end{center}
\caption{TH distributions in the model of ribosome traffic under 
PBC. Different curves correspond to different values of $\omega_{h}$, 
all for coverage density = 0.52 and $\omega_{a} = 2.5 s^{-1}$. 
}
\label{fig-riboTHwh}
\end{figure}

Under mean-field approximation, the master equations for the probabilities 
$P_{\mu}(i)$ are given by \cite{basu1}
\begin{equation} \label {3:0}
\frac{\partial{}P_{1}(i)}{\partial{}t} = \omega_{h2} P_{5}(i-1) Q(\underbar{i-1}|i-1+r) + \omega_{p} P_2(i) - \omega_{a} P_{1}(i)
\end{equation}
\begin{equation} \label {3:3}
\frac{\partial{}P_{2}(i)}{\partial{}t} = \omega_{a} P_{1}(i) - (\omega_{p} + \omega_{h1}) P_{2}(i)
\end{equation}
\begin{equation} \label {3:4}
\frac{\partial{}P_{3}(i)}{\partial{}t} = \omega_{h1} P_{2}(i) - k_{2} P_{3}(i)
\end{equation}
\begin{equation} \label {3:5}
\frac{\partial{}P_{4}(i)}{\partial{}t} = k_{2} P_{3}(i) - \omega_{g} P_{4}(i)
\end{equation}
\begin{equation} \label {3:6}
\frac{\partial{}P_{5}(i)}{\partial{}t} = \omega_{g} P_{4}(i) - \omega_{h2} P_{5}(i) Q(\underbar{i}|i+{r}) 
\end{equation}

In the steady state under PBC, the flux of the ribosomes is given by 
\begin{eqnarray} \label{3:7}
J = \frac{\omega_{h2} \rho (1- \rho {r})}{(1+\rho-\rho {r})+ \Omega_{h2}(1-\rho {r})}
\end{eqnarray}
where 
\begin{equation}
\Omega_{h2} = \omega_{h2}/k_{eff}.
\end{equation}
with
\begin{equation}
\frac{1}{k_{eff}} = \frac{1}{\omega_{g}}+\frac{1}{k_{2}}+\frac{1}{\omega_{h1}}+\frac{1}{\omega_{a}}+\frac{\omega_{p}}{\omega_{a}\omega_{h1}}
\end{equation}

TH distributions of the ribosomes in this model under PBC are plotted 
in figs.\ref{fig-riboTHd}, \ref{fig-riboTHwa}, \ref{fig-riboTHwh} for 
different sets of parameters \cite{agthesis}; the corresponding results 
under OBC have been reported elsewhere \cite{garairibo}.
The qualitative features of the TH distributions and their trend of 
variation with the parameters of the model of ribosome traffic are 
very similar to those observed in the preceeding section in the 
context of RNAP traffic. Neither $\tau_{min}$ nor the most probable 
TH depend on the coverage density. But, both $\tau_{min}$ and most 
probable TH decrease with increasing $\omega_{a}$ which is a measure 
of the abundance of the amino acid subunits of the growing protein chain.
A similar trend of variation of TH distribution is also observed with 
increasing $\omega_{h}$ which is a measure of the rate of ``fuel'' 
consumption (more precisely, rate of GTP hydrolysis). 

The width of the TH distribution in ribosome traffic can serve as 
a mesure of translational noise \cite{garairibo}, just as that in 
RNAP traffic has been considered in section \ref{sec-rnap} and 
in ref.\cite{tripathi} as a measure of transcriptional noise. 
A comparison between the trend of variation of this noise with 
our model parameters and the corresponding recent experimental 
observations has been reported elsewhere \cite{garairibo}.

\section{DH distributions in molecular motor traffic}
\label{sec-dhdist}

Because of the possibilities of attachments and detachments of the 
motors, the DH distribution is not a suitable quantity for 
characterizing KIF1A motor traffic. Therefore, in this section, 
we shall consider exclusively only traffic of RNAPs and ribosomes 
under PBC; the corresponding results under OBC are discussed in 
ref.\cite{ttthesis} and ref.\cite{agthesis}.

The DH distribution in the TASEP with hard rods, each of length $r$, 
was calculated analytically by Shaw et al.\cite{shaw03}. Suppose, 
$m$ denotes the DH. The DH-distribution $P(m)$ in this model under 
PBC is given by 
\cite{shaw03}  
\begin{equation}
P(m) = \biggl(\frac{\rho}{\rho_s}\biggr)\biggl(\frac{\rho_h}{\rho_s}\biggr)^{m}
\label{eq-DHformula}
\end{equation}
where $\rho_h = 1 - \rho_{cov}$ is the fraction of the system 
covered by the holes, and $\rho_s = \rho + \rho_h$. In the limit 
$r = 1$, $\rho_{cov} = \rho$, $\rho_s = 1$ and, hence, 
$P(m)$ reduces to the limiting form $P(m) = \rho (1-\rho)^{m}$ 
which is the well-known mean-field estimate of the DH distribution 
for TASEP \cite{dhchow,dhas}.

\begin{figure}[ht]
\begin{center}
\includegraphics[width=0.9\columnwidth]{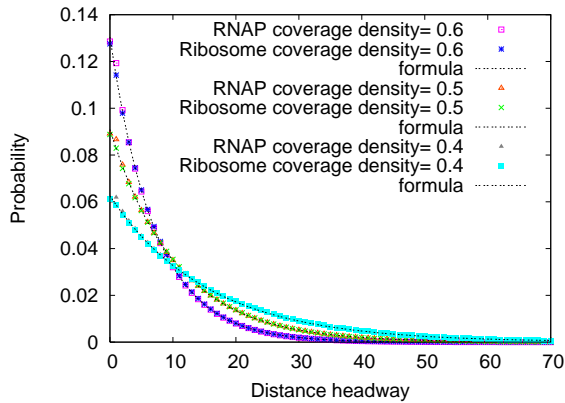}
\end{center}
\caption{DH distribution in RNAP traffic and ribosome traffic.
}
\label{fig-DHdist}
\end{figure}

In fig.\ref{fig-DHdist} we have plotted the DH distributions in the 
models of RNAP traffic and ribosome traffic under PBC where $r = 10$ 
for both the models. Interestingly, for any given coverage density, 
the DH distributions in both these models follow equation 
(\ref{eq-DHformula}) in spite of the differences in the 
mechano-chemical cycle of the individual motors. Thus, the DH 
distributions in these models of molecular motor traffic are determined
solely by the geometric parameters.

\section{Summary and conclusion}
\label{sec-conclusion}

In this paper we have modelled molecular motor transport on filamentary 
tracks from the perspective of physicists and traffic scientists. These 
models are essentially biologically motivated extensions of TASEP. To 
our knowledge, TASEP is the simplest model of non-equilibrium systems 
of interacting self-driven particles. Moreover, many extensions of TASEP 
have been used earlier to capture various phenomena observed in vehicular 
traffic. The interesting quantities which are used for characterizing 
traffic flow include fundamental diagram, distributions of distance- 
and time-headways. In this paper, for all the models of molecular motor 
traffic, we have not only summarized the earlier results on the fundamental 
diagrams, but also presented many new results on the distributions of 
DH and TH. The TH distribution in traffic of RNAPs and ribosomes yield 
novel measures of intrinsic noise in transcription and translation, 
respectively \cite{tripathi,garairibo}. Our modeling provide new insight 
into intracellular motor-driven processes by looking at these from a 
novel perspective.

\vspace{1cm}

\noindent {\bf Acknowledgements:} It is our great pleasure to thank 
all our collaborators for enjoyable collaborations. This work is 
supported (through DC) by a research grant from CSIR (India).



\end{document}